# To the problem of mathematical and physical modeling of the turbulent jets of mutually immiscible liquids like the oil and water

Ivan V. Kazachkov [1,2]

[1] *Dept of Information technology and data analysis, Nizhyn Gogol state university, Grafs'ka, 2, Ukraine, 16600, Tel: +380973191343; Email: ivan.kazachkov@energy.kth.se*

[2] *Dept of Energy Technology, Royal Institute of Technology, Sweden*

**Abstract**

Peculiarities of the turbulent two-phase and multiphase flows of the mutually immiscible liquids and averaged differential equations for their modeling are considered based on the approach, which was first developed and proposed by Prof. A.I. Nakorchevski as an alternative to a number of the well-known averaged multiphase dynamics equations. The main difference of the new method was in an averaging of the Navier-Stokes equations by phases and components in time instead of the widely spread spatial averaging in multiphase mechanics. What was more, introduction of the so-called function-indicator of the phases in a flow allowed recognizing the phases in their movement in a multiphase mixture, both theoretically and experimentally. For experimental study the special micro sensor was invented and created, which was successfully applied. In this paper, the method of Nakorchevski is described in application to some multiphase tasks, and a discussion is presented as concern to advantages of the method for study of turbulent two-phase flows of water-oil immiscible mixture, as well as many other similar mixtures, where it is important to know the peculiarities of the phases' movement and their mixing in a multiphase flow.

## 1. INTRODUCTION TO THE PROBLEM
### 1.1. The method of A.I. Nakorchevski for turbulent jets of immiscible liquids

The first attempt to develop a method, which naturally reflects the most important features of the flows of two or more immiscible liquids like a water and an oil was the one by Prof. A.I. Nakorchevski [1]. The characteristics of the mixture $a^l(t)$ (mass, velocity, impulse, etc.) of the corresponding characteristics of different phases $a_i^l(t)$ in a multiphase flow was proposed be expressed as follows

$$a^l(t) = \sum_{i=1}^{m} B_i(t) a_i^l(t), \qquad (1)$$

where $B_i(t)$ was introduced as co-called function-indicator determined as

$$B_i(t) = \begin{cases} 1, & \text{if } i-\text{phase occupies the elementary volume } \delta V \\ 0, & \text{if } i-\text{phase is outside the elementary volume } \delta V \end{cases}. \qquad (2)$$

With this approach, the analog of the Navier-Stokes equations in a boundary layer approximation was derived:

$$\frac{\partial}{\partial x}(y\rho_i B_i u_i) + \frac{\partial}{\partial y}(y\rho_i B_i v_i) = 0, \qquad \sum_{i=1}^{m} B_i = 1,$$

$$\rho_i B_i (u_i \frac{\partial u_i}{\partial x} + v_i \frac{\partial u_i}{\partial y}) = -\frac{dp}{dx} + \frac{1}{y}\frac{\partial}{\partial y}[yB_i \tau_i]_m, \qquad (3)$$

where is the sum by mute index *i*, which belong to a phase *i*. In the stationary equations of incompressible liquids (3) written in a cylindrical coordinate system are: $p$- pressure, $\rho$- density, $u,v$- the longitudinal and transversal velocity components, $\tau_i$ - turbulent stress for a phase *i*. Index *m* belongs to the values at the axis of the flow (symmetry axis). All values are averaged on the given interval by time. The function-indicator of a phase in multiphase flow may be considered as the mathematical expectation, in contrast to other multiphase approaches, which are based on introduction of the volumetric specific content of a phase in multiphase flow.

**Schematic representation of turbulent jet of two immiscible liquids**

The schematic representation of the turbulent two-phase flow of two immiscible liquids is given in Fig. 1:

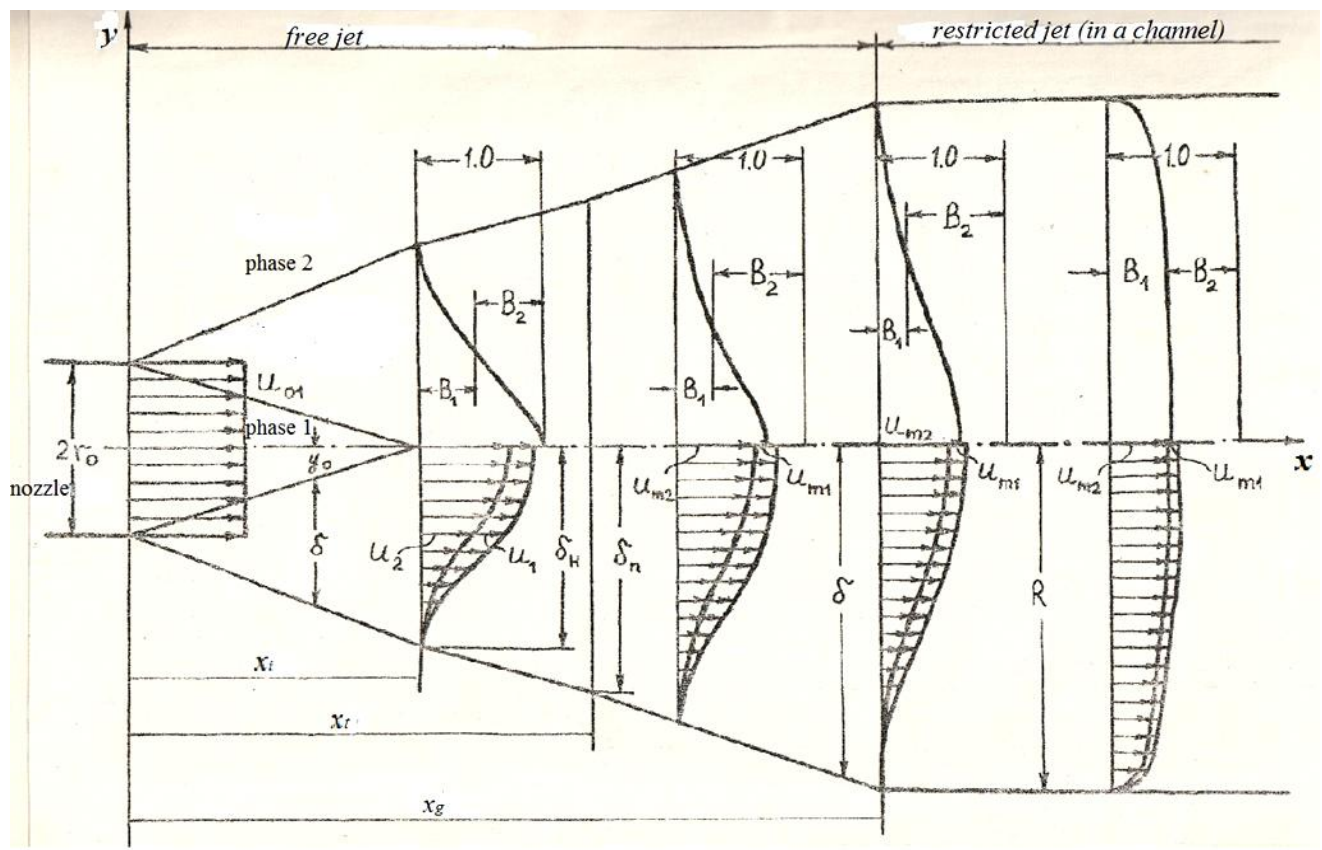

Fig. 1 The structural scheme of the two-phase jet

One liquid flow is going from the nozzle of the radius $r_0$ with velocity $u_{01}$ (the velocity profile is supposed simple uniform) in a surroundings occupied by other liquid (phase 2) being in a rest. The structure of a jet's mixing with surrounding liquid according to the Fig. 1 is simplified according to a traditional scheme [2]. First, the initial part of the length $x_i$ with the approximately linear boundaries for the conical surface (in cylindrical coordinates) of the internal potential core of a first phase and the external interface (conical too) are considered as the boundaries of the mixing area and the central potential core, correspondingly. The mixing turbulent zone between the above surfaces contains drops and fragments of the phases as far as immiscible liquids have behaviors like the separate phases, with their interfacial multiple surfaces interacting in all such locations (exchange of mass, impulse, energy among the phases). Close to the external interface there is a flow of a surrounding liquid with some drops of the first phase, while, in turn, close to the boundary of the potential core there is a flow of the first phase with the drops of a second phase. As far as we have mutually immiscible liquids, the mixing zone contains two phase mixture in a turbulent flow. After an initial part of the mixing zone when all first phase in a potential core is spent, the short transit area is preparing the ground part of the turbulent two-phase jet, where both phases are well mixed across all layer.

Normally for description of multiphase flows the spatial averaging of the differential equations of mass, impulse and energy conservation is performed using the concept of volumetric phase content [3-6], which does not fit so well to experimental study of the separate phase movement in a mixture as the approach proposed in [1], where the special experimental technology and micro sensor for some measurements in two-phase flows has been developed as well. In [4, 5] fundamentals and analysis of the different methods for modeling of the multiphase systems can be found. Actually all known methods of multiphase methods are well connected including the one in [1], and the parameters averaged by time can be easily transformed to the ones by [3-6].

## 1.2. The function-indicator of the phases and parameters of heterogeneous flow

The external interface of the mixing zone is determined zero longitudinal velocity of the second phase and transversal velocity of the first phase (the second phase is sucked from immovable surrounding into the mixing zone). The function-indicator of the first phase $B_1(t)$ is zero at the external interface because the first phase is absent in surroundings. Similar, the function-indicator $B_2(t)$ is zero on the interface of the potential core, boundary of the first phase going from the nozzle. In a first approach, an influence of the mass, viscous and capillary forces is neglected. With account of the above-mentioned, the boundary conditions are stated as follows [1]:

$$y=y_0, \ u_i=u_{0i}, \ v_i=0, \ \tau_i=0, \ B_1=1; \quad y=y_0+\delta, \ u_i=0, \ v_i=0, \ \tau_i=0, \ B_1=0. \quad (4)$$

The turbulent stress in the phase is stated by the "new" Prandtl's formula

$$\tau_i = \rho_i \kappa_i \delta u_{mi} \partial u_i / \partial y , \quad (5)$$

where $\kappa_i$ is the coefficient of turbulent mixing for $i$-th phase, $\delta$ is the width of the mixing layer.

## 1.3. The polynomial approximations of the flow parameters across the layer of a jet

The polynomial approximations for the velocity profiles and other functions in the turbulent mixing zone have been obtained based on the boundary conditions (4) in the form:

$$u_1 / u_{01} = 1 - 4\eta^3 + 3\eta^4, \quad (6)$$

$$u_2 / u_{02} = 1 - 6\eta^2 + 8\eta^3 - 3\eta^4, \quad (7)$$

$$B_1 = B_1^{(0)} = 1 - \eta^3 + 0.5\eta^2(1-\eta)h(x), \quad h \in [-6, 0],$$

$$B_1 = B_1^{(1)} = 1 - 4\eta^3 + 3\eta^4 + 0.5\eta^2(1-\eta)^2 h(x), \quad h \in [-12, -6],$$

$$B_1 = B_1^{(2)} = 1 - 10\eta^3 + 15\eta^4 - 6\eta^5 + 0.5\eta^2(1-\eta)^3 h(x), \quad h \in [-20, -12],$$

$$B_1 = B_1^{(3)} = 1 - 20\eta^3 + 45\eta^4 - 36\eta^5 + 10\eta^6 + 0.5\eta^2(1-\eta)^4 h(x), \quad h \in [-30, -20], \quad (8)$$

$$B_1 = B_1^{(4)} = 1 - 35\eta^3 + 105\eta^4 - 126\eta^5 + 70\eta^6 - 15\eta^7 + 0.5\eta^2(1-\eta)^5 h(x), \quad h \in [-42, -30],$$

$$B_1^{(5)} = 1 - 56\eta^3 + 210\eta^4 - 336\eta^5 + 280\eta^6 - 120\eta^7 + 21\eta^8 + 0.5\eta^2(1-\eta)^6 h(x), \quad h \in [-56, -42],$$

$$B_1^{(6)} = 1 - 84\eta^3 + 378\eta^4 - 756\eta^5 + 840\eta^6 - 540\eta^7 + 189\eta^8 - 29\eta^9 + 0.5\eta^2(1-\eta)^7 h(x), \quad h \in [-72, -56],$$

where $h(x)=\left(\partial^2 B_1 / \partial \eta^2\right)_{\eta=0}$ is an interesting function, which determines a transition of the piecewise continuous function-indicator $B_1^{(n)}$ to its next approximation, determined from the condition that the derivative by $\eta$ with respect to a point $\eta=1$ be equal to zero up to $(n+1)$-th and including order.

## 2. THE BASIC EQUATIONS FOR THE HETEROGENEOUS TWO-PHASE JET
### 2.1. Integral correlations for the initial part of a jet flow

Based on the above considered approximations the integral correlations have been derived for the two-phase turbulent jet on the initial part according to the structural scheme in Fig. 1 [1]:

$$u_{01}\left(r_0^2 - y_0^2\right) = 2\delta \int_0^1 B_1 u_1 (y_0 + \delta\eta) d\eta,$$

$$\rho_1 u_{01}^2 \left(r_0^2 - y_0^2\right) = 2\delta \int_0^1 \left(\rho_1 B_1 u_1^2 + \rho_2 B_2 u_2^2\right)(y_0 + \delta\eta) d\eta, \tag{9}$$

$$\rho_1 u_{01}\left(u_{01} - u_1^*\right) y_0 y_0' + \frac{d}{dx} \delta \int_0^{\eta^*} \sum_{j=1}^2 \rho_j B_j u_j^2 (y_0 + \delta\eta) d\eta - \sum_{j=1}^2 u_i^* \frac{d}{dx} \delta \int_0^{\eta^*} \rho_i B_i u_i (y_0 + \delta\eta) d\eta =$$

$$= \left(y_0 + \delta\eta^*\right) \sum_{j=1}^2 \rho_j B_j \kappa_j u_{0j} \frac{\partial u_j^*}{\partial \eta}, \qquad B_1 + B_2 = 1.$$

The first equation in (9) was got integrating by $y$ the mass conservation equation, the second and the third ones – integrating the impulse conservation for the total flow of a two-phase mixture for $y=y_0+\delta$ and $y=y^*$, respectively. The polynomial approximations for the functions $u_2$, $B_1$ on a ground part of the jet keep the same but for the function $u_1$ approximation is as follows

$$u_1 / u_{m1} = 1 - 3\eta^2 + 2\eta^3, \tag{10}$$

### 2.2. Integral correlations for the ground part of a jet flow

The integral correlations for the ground part of a jet obtained similarly to the above described initial part are as follows [1]:

$$2\int_0^\delta B_1 u_1 y\, dy = u_{01} r_0^2, \quad 2\sum_{j=1}^2 \int_0^\delta \rho_j B_j u_j^2 y\, dy = \rho_1 u_{01}^2 r_0^2, \tag{11}$$

$$\frac{d}{dx} \sum_{j=1}^2 \int_0^{y^*} \rho_j B_j u_j^2 y\, dy - \sum_{j=1}^2 u_j^* \frac{d}{dx} \int_0^{y^*} \rho_j B_j u_j y\, dy = y^* \sum_{j=1}^2 B_j^* \tau_j^*,$$

where the first is equation of the mass conservation for the first phase, the second and the third – the momentum conservation equations for the total and for the part of the cross section, respectively, according to the methodology [2]. And the momentum equation on the jet's axis ($y=0$) is used too:

$$\sum_{j=1}^2 \rho_j B_{mj} u_{mj} \frac{du_{mj}}{dx} = 2 \sum_{j=1}^2 \left[\frac{\partial}{\partial y}\left(B_j \tau_j\right)\right]_m. \tag{12}$$

The mathematical model including the ordinary differential equations (9), (11), (12) by longitudinal coordinate $x$ are implemented for analysis and numerical simulation on computer the basic features of the stationary turbulent two-phase jet of two immiscible liquids. The function-indicator $B_1$ shows how much is a presence of the first phase in a selected point of mixing zone, which can be directly compared to an experimental data by two-phase sensor. Therefore, a solution of the task may give both parameters of the flow together with their belonging to a particular phase.

**2.3. Dimensionless form of the outgoing equation array and the profiles of basic parameters**

The equation array (9) with the boundary conditions (4) is used for numerical simulation of the turbulent two-phase jet on its initial part. For this, the equations (9) are transformed to the following dimensionless form with the scales $r_0$, $\delta$, $u_{0i}$ for the longitudinal and transversal coordinates and velocity, respectively:

$$y_0^2 + 2\delta \sum_{j=1}^{2} y_0^{2-j} \delta^{j-1} a_j = 1, \quad y_0^2 + 2\delta \sum_{j=1}^{2} y_0^{2-j} \delta^{j-1} \left( a_{j+2} + i_0 b_{j+2} \right) = 1, \quad (13)$$

$$\left(1 - u_1^*\right) y_0 \frac{dy_0}{d\varsigma} + \frac{d}{d\varsigma} \delta \sum_{j=1}^{2} y_0^{2-j} \delta^{j-1} \left( a_{j+2}^* + i_0 b_{j+2}^* \right) - \frac{d}{d\varsigma} \delta \sum_{j=1}^{2} y_0^{2-j} \delta^{j-1} \left( a_j^* u_1^* + i_0 b_j^* u_2^* \right) = \left( y_0 + \delta \eta^* \right) =$$

$$= \left( y_0 + \delta \eta^* \right) \sum_{j=1}^{2} B_j^* \left( \frac{\partial u_j}{\partial \eta} \right)^* \left( i_0 \kappa_{21} \right)^{j-1}, \quad B_1 + B_2 = 1.$$

Here are:

$$\bar{y}_0 = y_0 / r_0, \ \bar{\delta} = \delta / r_0, \ \eta = (y - y_0) / \delta, \ \bar{x} = x / r_0, \ \varsigma = \kappa_1 \bar{x}, \ s_0 = u_{02} / u_{01}, \ i_0 = n s_0^2,$$

$$n = \rho_2 / \rho_1, \ \kappa_{21} = \kappa_2 / \kappa_1, \ a_i = a_{i1} + a_{i2} h, \ b_i = b_{i1} + b_{i2} h, \quad (14)$$

$$a_i = \int_0^1 B_1 \bar{u}_1 \eta^{j-1} d\eta, \ b_i = \int_0^1 B_2 \bar{u}_2 \eta^{j-1} d\eta \ (i=1, 2); \ a_i = \int_0^1 B_1 \bar{u}_1^2 \eta^{j-1} d\eta, \ b_i = \int_0^1 B_2 \bar{u}_2^2 \eta^{j-1} d\eta \ (i=3, 4); \ j=1, 2.$$

Except the above, for the dimensionless parameters we retain the same notations as for the dimensional ones. Only here in (14) it is stated for clarification of the dimensionless notations. The sliding factor $s_0$ is supposed to be constant and the values of the parameters at $\eta=\eta^*<1$ are signed with a star *. The system (13) must satisfy the following boundary conditions

$$\varsigma=0, \ y_0=1, \ \delta=0; \quad \varsigma=\varsigma_i, \ y_0=0, \ \delta=\delta_i; \quad (15)$$

where $\varsigma_i$, $\delta_i$ are the dimensionless length of a jet and its maximal radius (at the end of initial part).

By the computed from the above boundary problem (13), (15) functions $y_0(\varsigma)$, $\delta(\varsigma)$, $h(\varsigma)$ we can find all the other characteristics of the jet flow for the stated values of the main parameters of the model $i_0$, $\kappa_1$, $\kappa_2$. The first parameter is slightly indefinite due to difficulties with exact estimation of the phases' sliding, while the other two are known from the experimental studies but only for specific conditions. In general, for each specific case the, the coefficients of turbulent mixing $\kappa_1$, $\kappa_2$ may be different, and it's the problem to estimate them correctly. This is the main problem with validation of the mathematical model using comparison with the experimental data. The main advantage of this model is the possibility to have all characteristics of a flow together with their belonging to a particular phase through the functions $B_1$, $B_2$.

The transversal velocities' distributions, interface interactions, the coefficients of the volumetric $q$ and mass ejection $g$ and kinetic energy $e_i$ for the phases in a flow are computed as follows [1]:

$$\frac{B_1 v_1}{\kappa_1 u_{01}} = B_1 u_1 \frac{d}{d\varsigma} \left( y_0 + \delta \eta \right) - \frac{1}{y_0 + \delta \eta} \frac{d}{d\varsigma} \left( 0.5 y_0^2 + a_1 y_0 \delta + a_2 \delta^2 \right),$$

$$\frac{B_2 v_2}{\kappa_1 u_{02}} = B_2 u_2 \frac{d}{d\varsigma} \left( y_0 + \delta \eta \right) - \frac{1}{y_0 + \delta \eta} \frac{d}{d\varsigma} \left( b_1 y_0 \delta + b_2 \delta^2 \right),$$

$$R_{21} = -\frac{1}{(y_0 + \delta\eta)\delta} \frac{\partial u_1}{\partial \eta} \left\{ \frac{d}{d\varsigma}\left(0.5 y_0^2 + a_1 y_0 \delta + a_2 \delta^2\right) + \frac{\partial}{\partial \eta}\left[(y_0 + \delta\eta) B_1 \frac{\partial u_1}{\partial \eta}\right]\right\},$$

$$q = 2 s_0 \delta (b_1 y_0 + b_2 \delta), \qquad (16)$$

$$e_1 = y_0^2 + 2\int_0^1 B_1 \bar{u}_1^3 (y_0 + \delta\eta)\delta d\eta, \quad e_2 = 2 i_0 s_0 \int_0^1 B_2 \bar{u}_2^3 (y_0 + \delta\eta)\delta d\eta.$$

For the short transient part of the jet there are no developed substantiated scheme, therefore it is not under consideration here, just short transient and then the ground part is considered, where the method is well elaborated and supported with the experimental data [1, 2]. The dimensionless equation array for the ground part of the turbulent two-phase jet (11), (12) is the next

$$2 B_{m1} u_{m1} \delta^2 \sum_{j=1}^{2} \alpha_{1j} h^{j-1} = 1, \quad 2 B_{m1} u_{m1} \delta^2 \sum_{j=1}^{2}\left(\alpha_{2j} + i_0 \beta_{2j}\right) h^{j-1} + i_0 \beta_{20} = 1, \qquad (17)$$

$$\frac{d}{d\varsigma} u_{m1}^2 \delta^2 \left[B_{m1} \sum_{j=1}^{2}\left(\alpha_{2j}^* + i_0 \beta_{2j}^*\right) h^{j-1} + i_0 \beta_{20}^*\right] - u_{m1} \frac{d}{d\varsigma}\left(u_{m1}\delta^2\right)\left[B_{m1}\sum_{j=1}^{2}\left(\alpha_{1j}^* u_1^* + i_0 \beta_{1j}^* u_2^*\right)h^{j-1} + i_0 u_2^* \beta_{10}^*\right] =$$

$$= \eta^* \delta u_{m1}^2 \left[(1 - i_0 \kappa_{21}) B_{m1} \sum_{j=1}^{2} \left(\frac{\partial u_j}{\partial \eta}\right)^* \gamma_j^* h^{j-1} + i_0 \kappa_{21} \left(\frac{\partial u_2}{\partial \eta}\right)^*\right].$$

Here are:

$$\bar{x} = \frac{x - x_t}{r_0}, \quad \bar{u}_{mi} = \frac{u_{mi}}{u_{0i}}, \quad \bar{u}_i = \frac{u_i}{u_{mi}}, \quad i_0 = n s_0^2, \quad \bar{B}_2 = \frac{B_2}{B_{m1}}, \quad \bar{B}_1 = \frac{B_1}{B_{m1}} = \gamma_1 + \gamma_2 h, \quad \int_0^1 \bar{B}_1 \bar{u}_1^i \eta d\eta = \sum_{j=1}^{2} \alpha_{ij} h^{j-1},$$

$$\int_0^1 \bar{B}_2 \bar{u}_2^i \eta d\eta = \frac{\beta_{i0}}{B_{m1}} + \beta_{i1} + \beta_{i2} h \quad (i=1, 2). \qquad (18)$$

The same as previously we use these notations for dimensionless values only here, and we keep previous assignments for dimensionless parameters as for the dimension ones in all other equations. Star * means a value by $\eta = \eta^* < 1$, $x_t$ is the length of a transient part of a jet flow. It is assumed that $u_{m2} = s_0 u_{m1}$ ($s_0$=const), which means that sliding of the phases is preserved the same as for the initial part of a jet. The boundary condition for the equation array (17) are stated in a form

$$\varsigma = 0, \quad u_{m1}=1, \quad B_{m1}=1, \quad \delta = \delta_t; \quad \varsigma = \infty, \quad u_{m1}=0, \quad B_{m1}=0, \quad \delta = \infty; \qquad (19)$$

$\delta_t$ is a radius of the jet at the transient cross section.

## 3. SOLUTION OF THE BOUNDARY PROBLEM AND NUMERICAL SIMULATION
### 3.1. Basics of the turbulent heterogeneous jet on the initial part

Solution of the boundary problem (17), (19) allows obtaining the functions $u_{m1}(\varsigma)$, $B_{m1}(\varsigma)$, $\delta(\varsigma)$ and $h(\varsigma)$ by the modeling parameters $i_0$, $\kappa_{21}$. And similar to the initial part, here it is also available to compute distribution of all parameters of a flow.

Computational experiments reveal basic features of the flows by the initial parameters and the conditions stated. For the initial part of a jet, the velocity distributions for the phases ($u_1$, $u_2$) and function-indicators of the phases ($B_1$, $B_2$) are stated, as well as the values of parameters at the cross section of the nozzle. The solution can be done as follows. The system (13) contains two algebraic and one differential equation. From the algebraic equation array the functions $y_0(h)$, $\delta(h)$ are got, and then the differential equation is expressed in a standard form

$$dh/d\zeta = F(h(\zeta), i_0, \kappa_{21}), \qquad (20)$$

prepared for a numerical solution. Certainly the equation array (13) could be solved numerically in general but the way we applied is more comprehensive for understanding the basic features of the system, with as much as possible analytical expressions showing the explicit functions.

The range of the function's $h(\zeta)$ variation is determined by substitution of the boundary conditions in the functions $y_0(h)$, $\delta(h)$, so that we obtain the next:

$$y_0 = \frac{1}{\sqrt{1 + 2a_1 \frac{a_3 + i_0 b_3 - a_1}{a_2 - i_0 b_4 - a_4} + 2a_2 \left(\frac{a_3 + i_0 b_3 - a_1}{a_2 - i_0 b_4 - a_4}\right)^2}}, \qquad \delta = y_0 \frac{a_3 + i_0 b_3 - a_1}{a_2 - i_0 b_4 - a_4}; \qquad (21)$$

$$h_0 = h(0) = \frac{a_{11} - a_{31} - i_0 b_{31}}{a_{32} - a_{12} - i_0 b_{32}}, \qquad h_i = h(\varsigma_i) = \frac{a_{21} - a_{41} - i_0 b_{41}}{a_{42} - a_{22} - i_0 b_{42}}. \qquad (22)$$

$h_0(i_0)$, $h_i(i_0)$ computed from the (21), (22) are presented in Fig. 2.

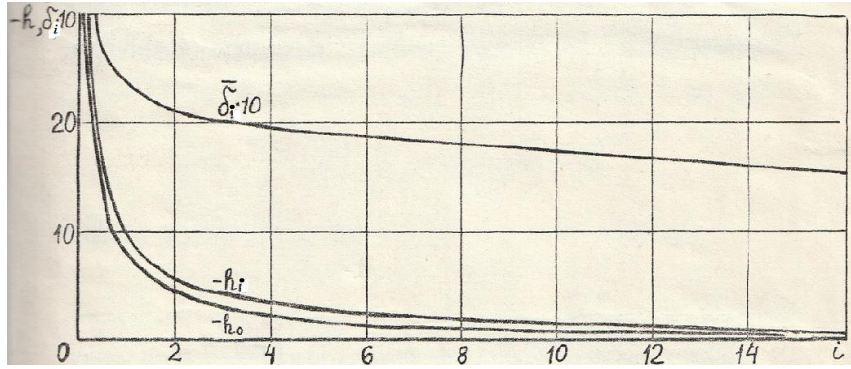

Fig. 2 Functions $h_i(i_0)$ and $\delta_i(i_0)$

As it is observed, the functions have the breaks at the transition points of the permanent characteristic function $B_1^{(n)}(\eta, h)$ from the one regional approximation to the other one (a first derivative has break at those points). It is impossible to get a common approximation for the function $B_1(\eta, h)$ satisfying the boundary conditions in all range by parameter $i_0$ (due to requirement of variation of function $B_1$ in the range from 0 to 1). Calculations have shown that there no substantial difference in the final results of the numerical solution at the points of the function $B_1(\eta, h)$ transformation, so that this approach is attainable.

Substitution of the values $h_i(i_0)$ into the equation (21) results in a relation for the radius of a jet at the end of the initial jet's part $\delta_i(i_0)$. The equations (21) determine the functions $y_0(h)$, $\delta(h)$ in the range $h \in [h_0, h_i]$, which are monotonous and close to the linear ones by $i_0$ over 5 being rapidly falling by $i_0$ below 1.

### 3.2. Basics of the turbulent heterogeneous jet on the ground part

The ground part of a jet is described by the equations (17) with the boundary conditions (19). From the first two equations yields

$$h = \frac{\alpha_{11} - (\alpha_{21} + i_0 \beta_{21})u_{m1} - 2i_0 \alpha_{11}\beta_{20}u_{m1}^2\delta^2}{-\alpha_{12} + (\alpha_{22} + i_0 \beta_{21})u_{m1} + 2i_0 \alpha_{12}\beta_{20}u_{m1}^2\delta^2}, \qquad B_{m1} = \frac{2i_0}{Z}\alpha_{12}\beta_{20} + \frac{\alpha_{22} + i_0\beta_{22}}{Zu_{m1}\delta^2} - \frac{\alpha_{12}}{Zu_{m1}^2\delta^2}, \qquad (23)$$

The condition $\zeta = \infty$, $B_{m1}=0$ leads the following: $\zeta = \infty$, $u_{m1}\delta^2 = \infty$, $u_{m1}\delta = const$, so that follows:

$$\lim_{\zeta \to \infty} \frac{\alpha_{12}}{Z}\left(2i_0\beta_{20} - \frac{1}{u_{m1}^2\delta^2}\right) = 0,$$

where from with account of $\alpha_{12} \neq 0$ yields

$$\lim_{\zeta \to \infty} u_{m1}^2 \delta^2 = \frac{1}{2i_0 \beta_{20}} \approx \begin{cases} 33, & i_0 = 0.3 \\ 10, & i_0 = 1.0 \\ 1.2, & i_0 = 8.0 \end{cases}, \quad h_\infty = \lim_{\zeta \to \infty} h \approx \begin{cases} -70, & i_0 = 0.3 \\ -31, & i_0 = 1.0 \\ -18, & i_0 = 8.0 \end{cases},$$

The value $h_\infty$ is outside the region of the function's $h$ variation determined by $h = h_t$, where the ground part of a jet starts. Therefore function $B_l$ is continuous for each value $i_0$ (its first derivative is piecewise continuous). It is changing its approximation with transformation from the one to another region by $h$. Thus, after simple transformations the outgoing equation array yields the system of two ordinary first order differential equations with the corresponding boundary conditions:

$$\frac{du_{m1}}{d\varsigma} = 2\frac{u_{m1}}{\delta}\left(u_1^"\right)_0 \frac{\left(u_1^"\right)_0 B_{m1} + i_0 \kappa_{21}(1-B_{m1})\left(u_2^"\right)_0}{\left(u_1^"\right)_0 \left[B_{m1} + i_0(1-B_{m1})\right]}, \quad (24)$$

$$\frac{d\delta}{d\varsigma} = \frac{1}{D_1 u_{m1}^2 \delta}\left[\left(M_1 + u_{m1}\delta^2 M_2 + \frac{M_3}{u_{m1}}\right)f_1 + 0.75\left(\frac{N_3}{\delta} + \frac{N_2 u_{m1}}{\delta} + N_1 u_{m1}^2 \delta\right)\right];$$

$$\varsigma = 0, \quad u_{m1} = 1, \quad \delta = \delta_i. \quad (25)$$

Here are:

$$M_2 = \frac{2i_0}{Z}\beta_{20}\left\{2\left[\alpha_{12}\left(\alpha_{21}^* + i_0 \beta_{21}^*\right) - \left(\alpha_{22}^* + i_0 \beta_{22}^*\right)\alpha_{11}\right] + \left[\alpha_{11}\left(\alpha_{12}^* u_1^* + i_0 \beta_{12}^* u_2^*\right) - \alpha_{12}\left(\alpha_{11}^* u_1^* + i_0 \beta_{12}^* u_2^*\right)\right]\right\} +$$

$$i_0\left(2\beta_{20}^* - \beta_{10}^* u_2^*\right), \quad f = \frac{du_{m1}}{d\varsigma}, \quad M_1 = \frac{1}{Z}\left[\left(\alpha_{21}^* + i_0 \beta_{21}^*\right)\left(\alpha_{22} + i_0 \beta_{22}\right) - \left(\alpha_{22}^* + i_0 \beta_{22}^*\right)\left(\alpha_{21} + i_0 \beta_{21}\right)\right],$$

$$M_3 = \frac{1}{Z}\left[\alpha_{11}\left(\alpha_{12}^* u_1^* + i_0 \beta_{12}^* u_2^*\right) - \alpha_{12}\left(\alpha_{11}^* u_1^* + i_0 \beta_{11}^* u_2^*\right)\right], \quad N_3 = \frac{1}{Z}\left(\alpha_{12}\gamma_1 - \alpha_{11}\gamma_2\right)\left(i_0 \kappa_{21} - 1\right),$$

$$D_1 = \frac{4i_0}{Z}\beta_{20}\left[\alpha_{11}\left(\alpha_{22}^* + i_0 \beta_{22}^*\right) - \alpha_{12}\left(\alpha_{21}^* + i_0 \beta_{21}^*\right) - \alpha_{11}\left(\alpha_{12}^* u_1^* + i_0 \beta_{12}^* u_2^*\right) + \alpha_{12}\left(\alpha_{11}^* u_1^* + i_0 \beta_{11}^* u_2^*\right)\right] +$$

$$+2i_0\left(\beta_{10}^* u_2^* - \beta_{20}^*\right), \quad N_1 = \frac{2i_0}{Z}\beta_{20}\left(\alpha_{12}\gamma_1 - \alpha_{11}\gamma_2\right)i_0\kappa_{21}\left[1 - \frac{2i_0}{Z}\beta_{20}\left(\alpha_{12}\gamma_1 - \alpha_{11}\gamma_2\right)\right],$$

$$N_2 = \frac{1}{Z}\left\{i_0\kappa_{21}\left[\gamma_2\left(\alpha_{12} + i_0 \beta_{21}\right) - \gamma_1\left(\alpha_{22} + i_0 \beta_{22}\right)\right] + \gamma_1\left(\alpha_{22} + i_0 \beta_{22}\right) - \gamma_2\left(\alpha_{21} + i_0 \beta_{21}\right)\right\}.$$

### 3.3. The solution procedure

The boundary problem (24), (25) was solved numerically with a control of the value $h$ and automatic transforming of the approximation $B_l$. The functions $u_{m1}(\zeta)$, $B_{m1}(\zeta)$, $\delta(\zeta)=0$, and $h(\zeta)$ are computer for a range of parameters $i_0$, $\kappa_{21}$. The control parameters are kept in the computer program as follows: $B_{m1}<1$, $u_{m1}<1$, $dB_{m1}/d\zeta < 0$, $d\delta/d\zeta > 0$. An analysis of the numerical simulations has shown that by the boundary conditions (25) the solution is correct only for the restricted regions by a ratio of turbulent mixing coefficients $\kappa_{21}$ (specific for each value of the parameter $i_0$). It's an interesting feature that a ration of the turbulent mixing coefficients of the phase cannot be arbitrary, which seems to be physically reasonable. For the average value of the $\kappa_{21}^{av}$ and the variation range $\Delta\kappa_{21}$ the following approximations were obtained:

$$\kappa_{21}^{av} = 0.2/i_0, \quad \Delta\kappa_{21} = \pm 0.02 i_0.$$

The ratio of the coefficients of turbulent mixing is hyperbolically falling down with increase of a density ratio: the higher is density of an ejected liquid, the lower is its mixing coefficient comparing to a first liquid. But possible interval of a ratio can grow with increase of the density ratio. It is understandable as heavy liquid loses its ability for intensive mixing with the other liquid.

Interesting feature was revealed about the influence of parameters $i_0$, $\kappa_{21}$ on solution of the problem. The radius of a jet and velocities of phases on an axis practically don't depend on $\kappa_{21}$ being totally determined by the value $i_0$, while the functions $B_{m1}$ and $h$ strictly depend on $\kappa_{21}$. Thus, the turbulent mixing influences mostly the internal structure of a flow, phases' distribution, and velocities depend on density ratio of the phases (internal structure has little influence on it). The velocity distribution, in a turn, determines the radius of a mixing zone because it changes with falling of the velocity according to the mass and momentum conservation equations.

Due to absence of the proven methodology for calculation of the transient part of a jet, the ground part of a turbulent two-phase jet is proposed for investigation in a following way, independent of the limitations by $\kappa_{21}$. As far as two functions, $B_{m1}$ and $h$, determine the function $B_1$, and an influence of $B_{m1}$ is stronger than $h$, we can assume $h = h_i =$ const for the ground part of a jet, so that the function $B_{m1}$ be correcting a possible inaccuracy of it. This assumption over determines the task, therefore we can have possibility for calculation of the same characteristic twice, independently. We choose a radius of the mixing zone $\delta$. Controlling a ration of $\delta_1$ and $\delta_2$ obtained from solution of two independent equations we can decide upon attainability of this our assumption, as well as total inaccuracy of the model. Then equations (24) yield:

$$B_{m1} = \beta_{20} \frac{u_{m1} i_0}{(\alpha_{11} + \alpha_{12} h_t) - \left[(\alpha_{21} + i_0 \beta_{21}) + (\alpha_{22} + i_0 \beta_{22}) h_t\right] u_{m1}}, \quad \delta = \frac{1}{\sqrt{2 B_{m1} u_{m1} (\alpha_{11} + \alpha_{12} h_t)}}. \quad (26)$$

Then the equation array (24), (26) with the boundary conditions (25) is solved as follows. From the equations (26) and the first equation of the system (24), the functions $u_{m1}, B_{m1}$ and $\delta_{=}\delta_1$ are obtained depending on the longitudinal coordinate $\zeta$ and parameters $i_0$, $\kappa_{21}$. The other radius of the turbulent zone $\delta_{=}\delta_2$ is obtained from the second equation of the system (24) by $\eta = \eta^* = 0.5$.

### 3.4. The results of computer simulations

The results of computations of the values $h_t(i_0)$, $\delta_t(i_0)$ and distribution of the velocity, function-indicator of a phase and the radius of the turbulent mixing zone are given in the Figs 3, 4:

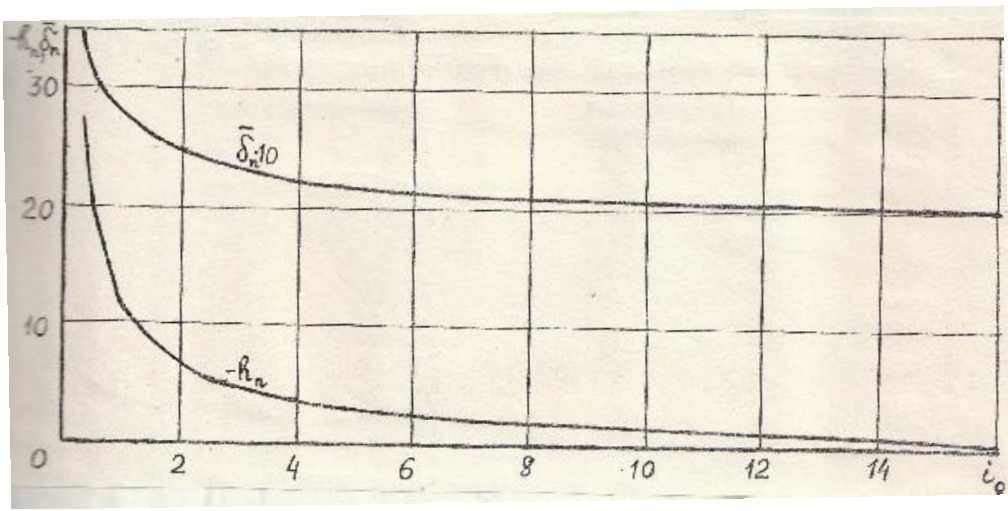

Fig. 3 Radius of a jet $\delta_t$ and function $h_t$ depending on parameter $i_0$ (density ratio)

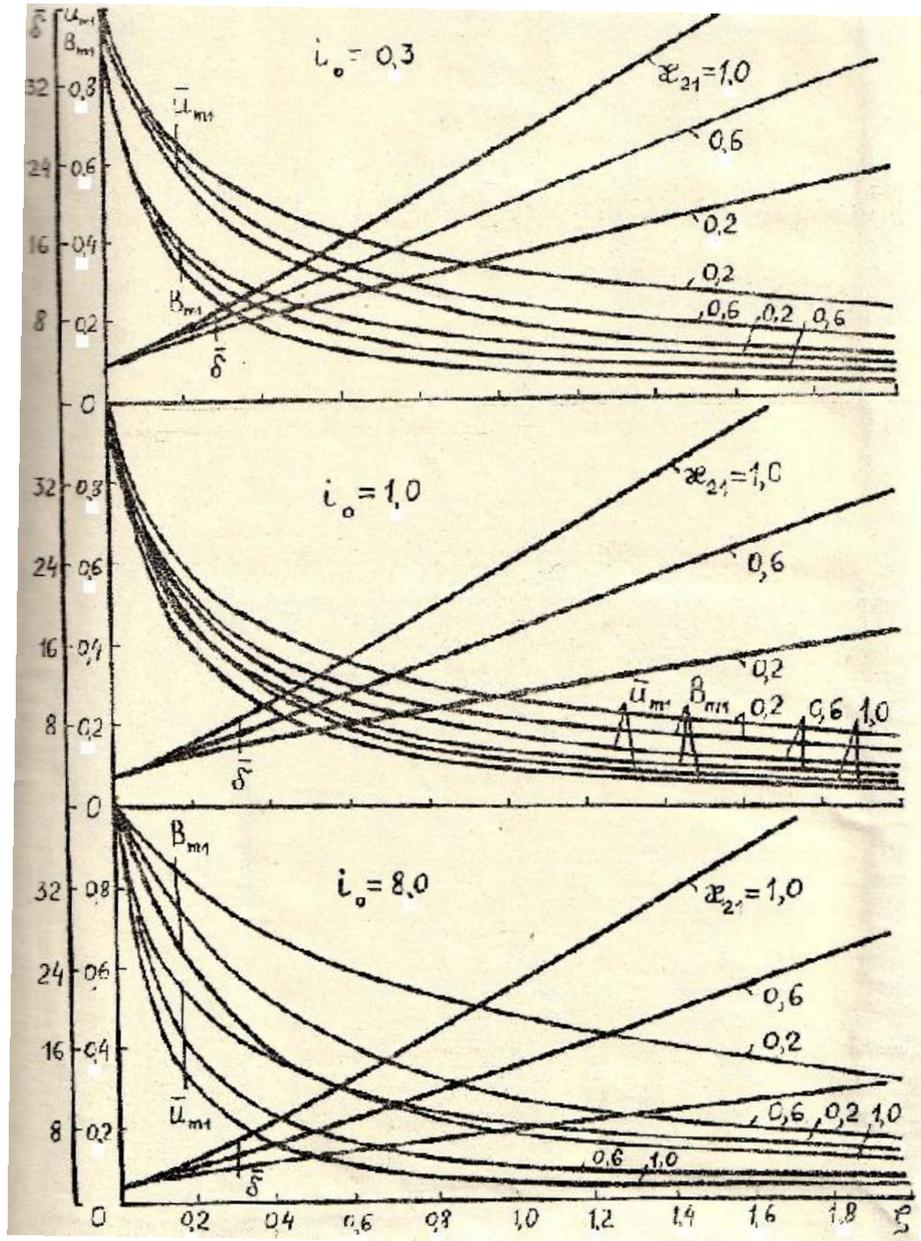

Fig. 4 Axial velocity, function-indicator and radius of the jet along the axis of a jet

In the next calculations, the value $\delta = \delta_1$ is accepted as a radius of a jet. Some of the computer simulations and the FORTRAN programs by the methodology described here first done by us were presented in [7-11]. Then it was further developed by Prof. Nakorchevki with his co-workers, as well as by other researchers in a number of papers, e.g. [12, 13].

The equation (20) represented in a form

$$dh/d\zeta = F, \qquad (27)$$

after substitution of the computed functions $y_0(h)$ $\delta(h)$, was solved in a following way. Function FM=1/F was computed in the range of parameters: $i_0 \in [0.2, 16]$, $\kappa_{21} \in [0, 5]$, $h \in [-20, 0]$. Its approximation was found as FM=$A_1' + A_2' h + A_3' h^2$, where $A_i'(i_0, \kappa_{21})$ - computed function with accuracy no less than 0.2%. The maximal total inaccuracy of computation after integration of (27) estimated by the Caushy-Bunyakovski inequality satisfied the condition $D_{max}<\Delta_{max}(\Delta h)_{max}$. As far as the Fig. 2 shows $(\Delta h)_{max}<4$, it was $D_{max}<0.8\%$. It is inside the attainable range of inaccuracy for the integral methods of the turbulent jets and boundary layers.
Integration of the equation (27) with account of the above-mentioned results

$$\zeta = A_0 + A_1 h + A_2 h^2 + A_3 h^3, \qquad (28)$$

where $A_0 = -A_1 h_0 - A_2 h_0^2 - A_3 h_0^3$, $A_i = A_i'/i$, $i=1,2,3$. Thus, $\zeta_i = A_0 + A_1 h_i + A_2 h^2_i + A_3 h^3_i$. Computation by this formula is presented in Fig. 5:

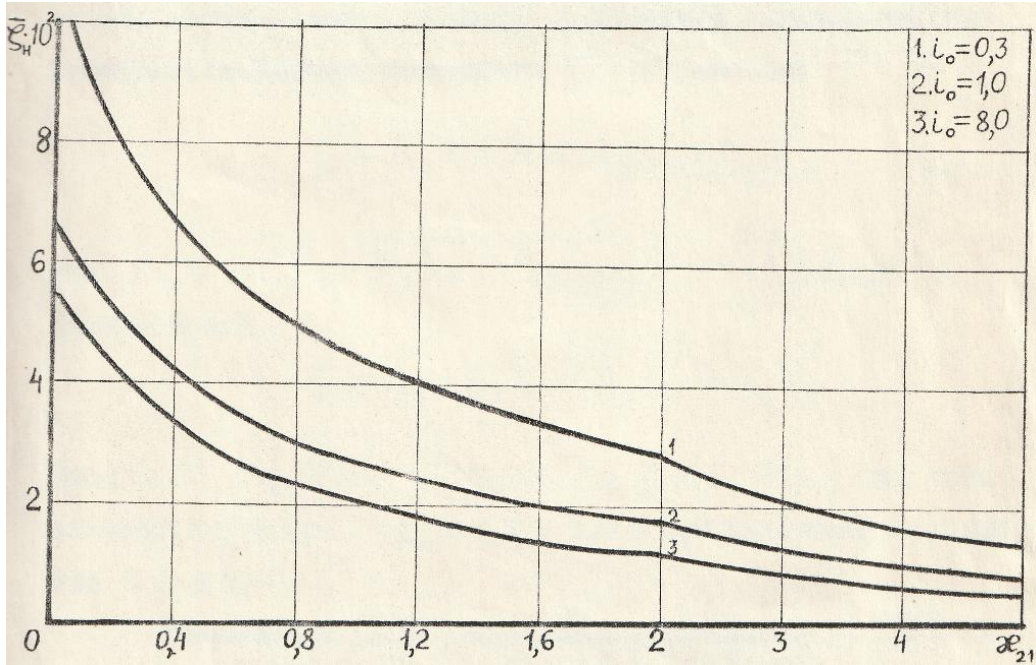

Fig. 5 The length of the initial part of a jet depending on the turbulent mixing ratio $\kappa_{21}$

where from follows that the length of the initial part of a jet strictly depends on a ratio of the turbulent mixing coefficients $\kappa_{21}$: the higher is turbulent mixing in a second phase (ejected from the surrounding medium), the longer is an initial part of a jet. Strong mixing in a first phase coming from the nozzle means physically that it keeps parameters of the first phase for a long time. With a strong mixing coefficient of an injected phase, an initial part of a jet is short independent of density ration of the phases in a wide range from $i_0=0.3$ till $i_0=8$. If the mixing coefficients in a phases are of the same order or coefficient in an ejected liquid is less than in a liquid going from the nozzle, the initial part of the jet is long (up to 5-10 times bigger than a radius of a nozzle). And the denser is first liquid, the long is an initial part of a jet.

The functions $A_i(i_0, \kappa_{21})$, $\zeta_i(i_0, \kappa_{21})$, $i=1,2,3$, for 3 different values of $i_0$ are presented in Figs 6-8. The correlation (28) allows computing functions $y_0(h)$ $\delta(h)$, and current radius of a jet $r=y_0(h)+\delta(h)$, shown in Fig. 7 (1- $\kappa_{21}=0.2$, 2- $\kappa_{21}=1$, 3- $\kappa_{21}=5.0$). The other characteristics of the turbulent jet computed by the main parameters obtained, e.g.

$$\bar{\tau} = \frac{\tau}{\rho_1 \kappa_1 u_{01}^2} = 12\eta(\eta-1)\left[\eta B_1 + i_0 \kappa_{21}(1-\eta) B_2\right], \quad \frac{B_i v_i}{u_{0i}}, \quad \frac{v}{u_{01}} = \frac{B_1 v_1}{u_{01}} + s_0 \frac{B_2 v_2}{u_{02}}, \quad \langle\overline{\rho u^2}\rangle = \frac{\langle\rho u^2\rangle}{\rho_1 u_{01}^2} = B_1 \bar{u}_1^2 + i_0 B_2 \bar{u}_2^2.$$

They are presented in Figs 9-12. Parameters of the turbulent two-phase jet, both analytical and numerical solutions obtained, have been compared against experimental data for the oil-water immiscible liquids (see Fig. 13).

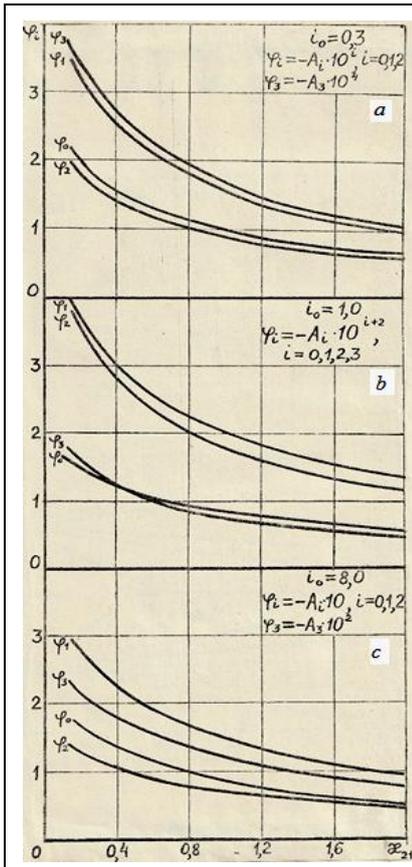

Fig. 6 The functions $A_i(i_0, \kappa_{21})$: against $\kappa_{21}$ for 3 values of $i_0$

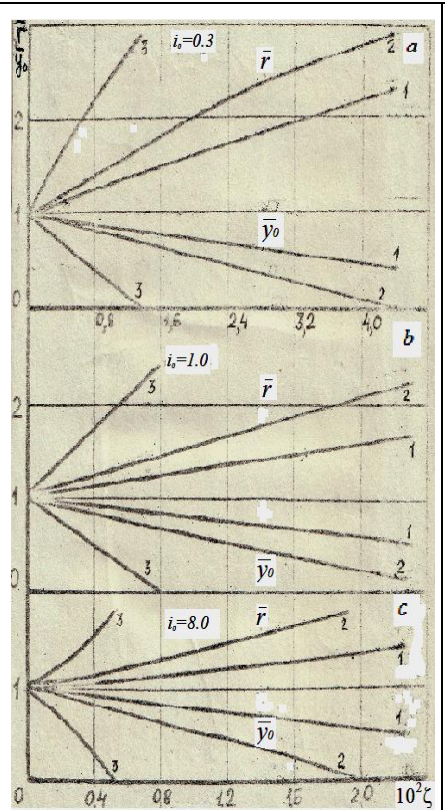

Fig. 7 Radius of potential core and mixing zone of two-phase jet

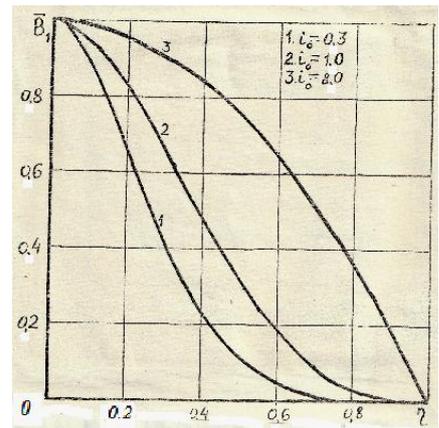

Fig. 8 Function-indicator of the first phase across mixing zone of a jet

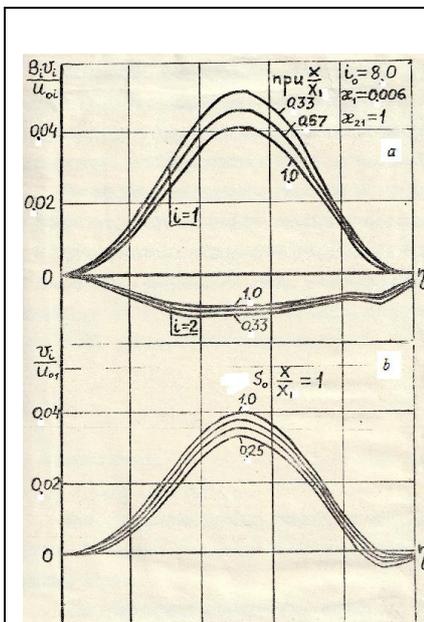

Fig. 9 $A_i(i_0, \kappa_{21})$: against $\kappa_{21}$ and $i_0$

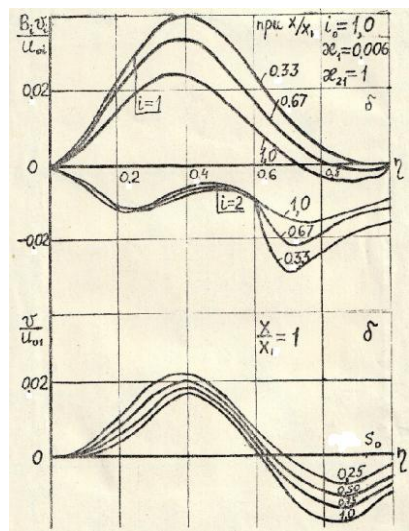

Fig. 10 Radius of potential core and mixing zone of two-phase jet

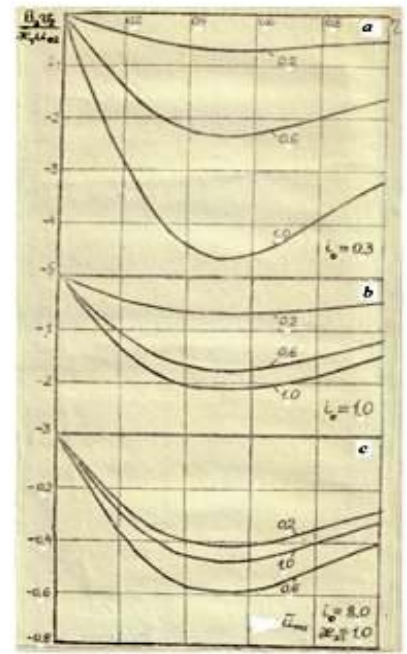

Fig. 11 Function-indicator of the first phase across two-phase jet

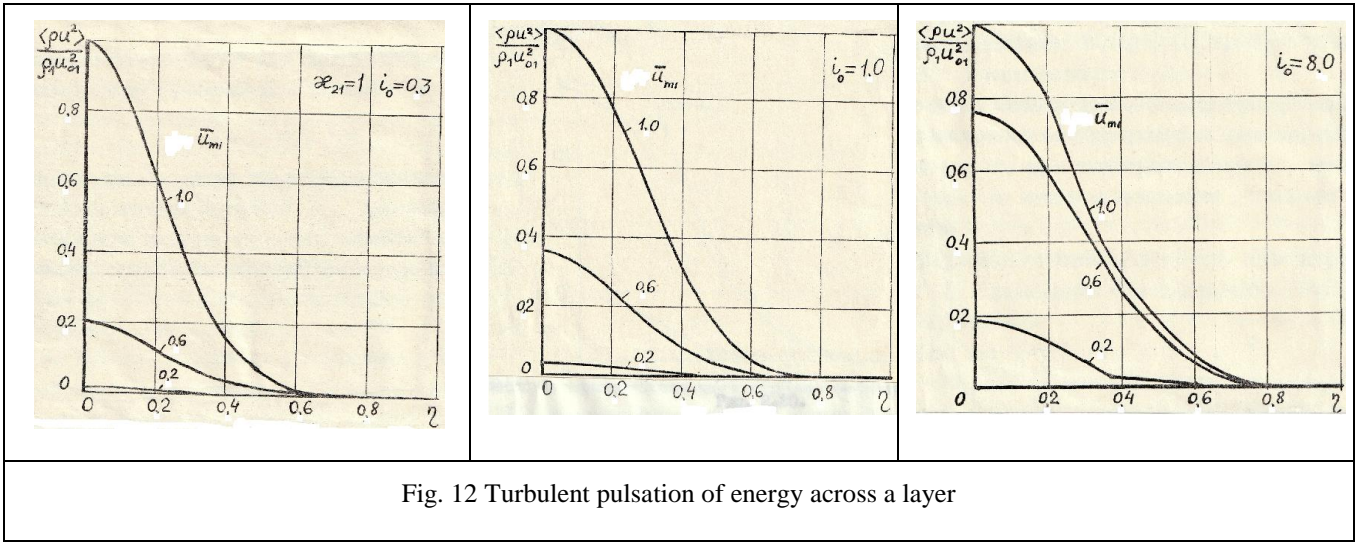

Fig. 12 Turbulent pulsation of energy across a layer

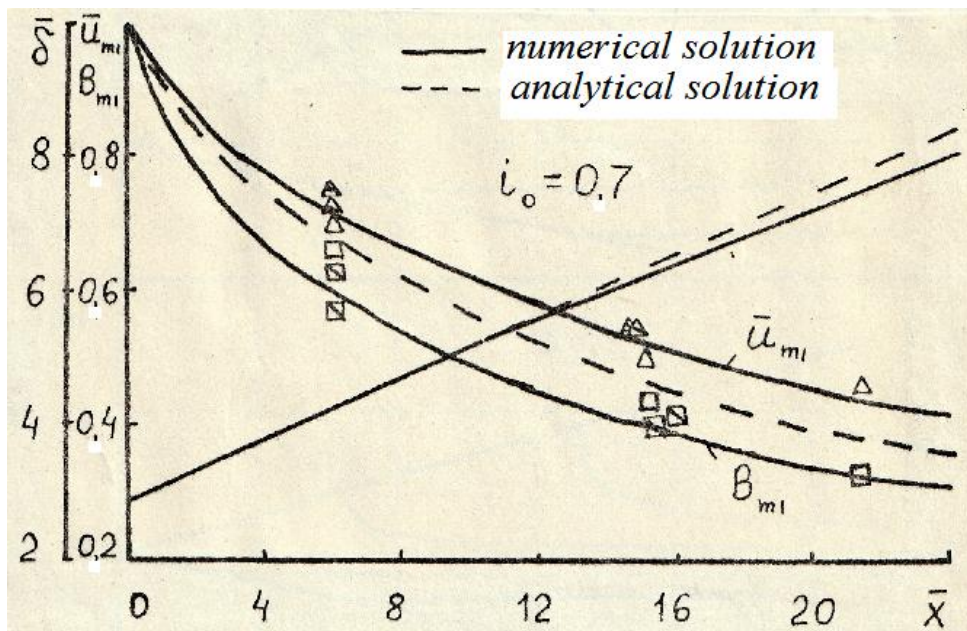

Fig. 13 Parameters of turbulent two-phase jet against experimental data for oil-water immiscible liquids

The experimental study was performed using the special two-phase micro sensor, which was developed and constructed by Prof. A.I. Nakorchevski and his PhD student V.A. Chernov. The sensor was prepared by micro miniaturist Docent NTUU "KPI" V. Tinyakov.

The stream lines in a turbulent two-phase jet of the immiscible liquids on its initial part, calculated by the parameters obtained in the numerical simulations, are presented in Fig. 14 for the three different values of the parameter $i_0$=0.3; 1.0; 8.0, and for the determined values of the turbulent mixing coefficients 0.006, ($\kappa_{21}$ =1). It is clearly observed from the results obtained that for all density ratios of the phases, the second phase (ejected liquid) does not penetrate into the mixing layer more than a half of it, except the high density ratio ($i_0$=8.0) when mixing of the liquids is really intensive across the all layer.

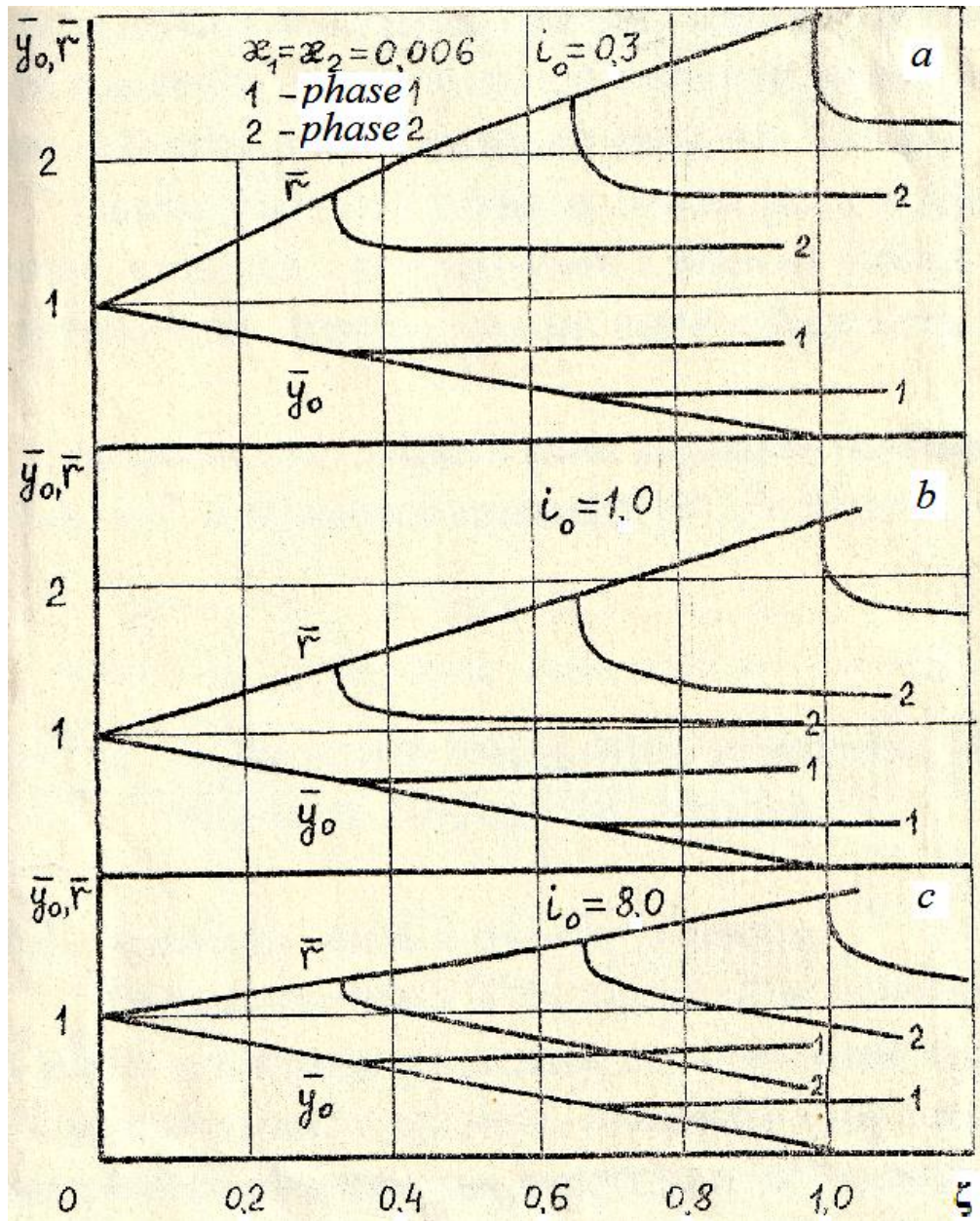

Fig. 14 The stream lines in a turbulent two-phase of the immiscible liquids

**Acknowledgements**

The author devotes this article to the blessed memory of Professor's Alfred I. Nakorchevski, who was the first his mentor in scientific work. And the author wishes to acknowledge Professor Torsten Henry Fransson for the possibility to work at the Department of Energy Technology, Royal Institute of Technology (KTH), during many years since 2001. Without his support, I would not have the possibility to work in science due to the bad economic situation and lack of financing of the research in Ukraine. He helped me to survive as a scientist and obtain highly valuable new experience working in the international scientific environment of the KTH.